\documentclass[aps,nofootinbib,prl,superscriptaddress,tightenlines,notitlepage,twocolumn,showpacs,floatfix]{revtex4-1}
\usepackage[colorlinks]{hyperref}
\usepackage{tabularx}
\usepackage{graphicx}
\usepackage{amssymb,bm,tensor}
\usepackage{textcomp}
\usepackage{xcolor}
\usepackage{amsmath}
\usepackage[varg]{txfonts}
\usepackage{enumerate}
\usepackage{mathtools}
\usepackage[normalem]{ulem}
\usepackage{bm} % bold math
\usepackage[OT1]{fontenc}
\usepackage{aas_macros}

\newcommand{\rmd}{{\rm d}}
\newcommand{\MPIfR}{\affiliation{Max-Planck-Institut f\"ur Radioastronomie, Auf
dem H\"ugel 69, D-53121 Bonn, Germany}}
\newcommand{\Manchester}{\affiliation{Jodrell Bank Centre for Astrophysics, The
University of Manchester, M13 9PL, United Kingdom}}

\begin{document}

\title{Testing the universality of free fall towards dark matter with radio
pulsars}
\date{\today}
\author{Lijing Shao}\email{lshao@mpifr-bonn.mpg.de}\MPIfR
\author{Norbert Wex}\MPIfR
\author{Michael Kramer}\MPIfR\Manchester

\begin{abstract}
  The violation of the weak equivalence principle (EP) in the gravitational
  field of the Earth, described by the E\"otv\"os parameter $\eta_\oplus$, was
  recently constrained to the level $\left|\eta_\oplus\right| \lesssim
  10^{-14}$ by the MICROSCOPE space mission. The E\"otv\"os parameter
  $\eta_{\rm DM}$, pertaining to the differential couplings of dark matter (DM)
  and ordinary matter, was only tested to the level $\left| \eta_{\rm DM}
  \right| \lesssim 10^{-5}$ by the E\"ot-Wash group and lunar laser ranging.
  This test is limited by the EP-violating {\it driving force} in the Solar
  neighborhood that is determined by the Galactic distribution of DM.  Here we
  propose a novel celestial experiment using the orbital dynamics from radio
  timing of binary pulsars, and obtain a competing limit on $\eta_{\rm DM}$
  from a neutron star (NS) -- white dwarf (WD) system, PSR~J1713+0747.  The
  result benefits from the large material difference between the NS and the WD,
  and the large gravitational binding energy of the NS. If we can discover a
  binary pulsar within $\sim$\,10\,parsecs of the Galactic center, where the
  driving force is much larger in the expected DM spike, precision timing will
  improve the test of the universality of free fall towards DM and constrain
  various proposed couplings of DM to the Standard Model by several orders of
  magnitude.  Such a test probes the hypothesis that gravity is the only
  long-range interaction between DM and ordinary matter.
\end{abstract}
\pacs{95.30.Sf, 95.35.+d, 97.60.Gb, 98.35.Jk}
% 95.30.Sf  -- Gravitation (astrophysics)
% 95.35.+d  -- dark matter
% 97.60.Gb  -- pulsars
% 98.35.Jk  -- Galactic center

\maketitle

%% === main body of the paper ===

%---------------------------------------------------------------------
\paragraph{Introduction.}
\label{sec:intro}
%---------------------------------------------------------------------

In the opening paragraph of {\it Philosophi\ae{} Naturalis Principia
Mathematica}, Newton studied carefully the equivalence between {\it mass} and
{\it weight}, which is later known as the equivalence principle (EP). It lies
at the heart of Newtonian gravity, as well as Einstein's general relativity
(GR)~\cite{Will:1993ns, Adelberger:2009zz, Will:2014kxa, Nobili:2017cxu}.  As
emphasized by various authors~\cite{Fock:1964, Damour:2012rc, Nobili:2017cxu},
EP should be treated as a heuristic concept, instead of a {\it principle}.
Experimental examination of EP started with pendulums by Galileo, Newton,
Bessel, Potter {\it et al.}~\cite{Nobili:2017cxu}, and flourished with torsion
balances by E\"otv\"os, Dicke, Braginsky, Adelberger {\it et
al.}~\cite{Adelberger:2009zz}.  Recently, no violation was detected between
titanium and platinum alloys from the first result of the MICROSCOPE satellite
to the level $\left| \eta_\oplus^{\left({\rm Ti,Pt}\right)} \right| \lesssim
10^{-14}$~\cite{Touboul:2017grn} where the E\"otv\"os parameter (with subscript
denoting the attractor),
%---
\begin{equation}\label{eq:acc:eta}
  \eta_\oplus^{\left(A,B\right)} \equiv \frac{a_A - a_B}{\frac{1}{2}\left(a_A +
  a_B\right)} \,,
\end{equation}
%---
describes the difference in the acceleration of test bodies $A$ and $B$ in the
gravitational field of the Earth; in the following, we call the measurement of
the numerator {\it precision}, and the denominator {\it driving force}.  The
MICROSCOPE result surpasses the limits, by a factor of ten, from the E\"ot-Wash
group~\cite{Adelberger:2009zz, Wagner:2012ui}. On the other hand, the
E\"otv\"os parameter towards the Sun was constrained to be $\left| \eta_\odot
\right| \lesssim 10^{-13}$ by lunar laser ranging (LLR)~\cite{Williams:2012nc}
and E\"ot-Wash group~\cite{Wagner:2012ui}.  The physical distinction between
$\eta_\oplus$ and $\eta_\odot$ is necessary because in the analysis the {\it
driving forces} are produced by different composition of the attractor, mostly
hydrogen ($\sim91.2\%$) and helium ($\sim8.7\%$) for the Sun, and iron
($\sim32.1\%$), oxygen ($\sim30.1\%$), silicon ($\sim15.1\%$), and magnesium
($\sim13.9\%$) for the Earth. If EP violation is caused by a long-range force
mediated by a new massless (or ultralight) field, $\eta_\oplus$ and
$\eta_\odot$ probe different aspects of its couplings between the attractor and
test masses~\cite{Adelberger:2009zz}. MICROSCOPE has no gain in the driving
force from the Sun, thus it does not improve the limit on $\eta_\odot$.  The
composition of two test masses is also important. It is related to the
couplings between the force-mediating field and the force receivers. In this
regard to the gravitational energy, a branch of well-motivated studies use
self-gravitating bodies to probe the strong EP~\cite{Nordtvedt:1968qs,
Will:1993ns, Will:2014kxa} with LLR~\cite{Williams:2012nc} and pulsar timing
experiments~\cite{Damour:1991rq, Wex:1999bx, Stairs:2005hu, Gonzalez:2011kt,
Zhu:2015mdo, Wex:2014nva, Shao:2016ezh, Zhu:2018etc}.

\citet{Stubbs:1993xk} was the first to point out that, the E\"ot-Wash searches
for EP violation also put limits on the E\"otv\"os parameter $\eta_{\rm DM}$
when the dark matter (DM) acts as the attractor. It could originate from
differential couplings between DM and ordinary matter~\cite{Adelberger:2009zz,
Graham:2015ifn}. With its actual ability in measuring differential acceleration
worse than the E\"ot-Wash group, MICROSCOPE  benefits from a larger driving
force, $7.9\,{\rm m\,s}^{-2}$ at 710\,km altitude versus
$1.68\times10^{-2}\,{\rm m\,s}^{-2}$ for the E\"ot-Wash
laboratory~\cite{Wagner:2012ui, Touboul:2017grn}. However, because of a much
smaller driving force in the Solar neighborhood from the DM, $a_{\rm DM} \simeq
10^{-10}\,{\rm m\,s}^{-2}$~\cite{2017MNRAS.465...76M}, previous studies were
only able to constrain $\left|\eta_{\rm DM}\right| \lesssim
10^{-5}$~\cite{Smith:1993un, Su:1994gu, 1995PhRvD..51.3135S,
1994ApJ...437..529N, 1995A&A...293L..73N, Anderson:1995df, Williams:2005rv,
Wagner:2012ui}.  

In this {\it Letter} we demonstrate that the current limit in testing EP from
pulsars~\cite{Damour:1991rq, Stairs:2005hu, Freire:2012nb, Wex:2014nva,
Shao:2014wja, Shao:2016ezh, Zhu:2018etc} is already approaching the best
available constraint on $\eta_{\rm DM}$.  Considering (i) the neutron-rich
composition of pulsars, and (ii) their significant amount of gravitational
binding energy, it is advantageous to translate the pulsar limit  into DM's
differential couplings between protons and neutrons~\cite{Adelberger:2009zz,
Wagner:2012ui}.  While all other tests are limited to the Solar neighborhood,
pulsar surveys towards the Galactic center (GC)~\cite{2010ApJ...715..939M,
Wharton:2011dv, Eatough:2015jka, DeLaurentis:2017dny} might find suitable
pulsars in the future with much larger driving forces from the Galactic DM
distribution (in particular the expected DM spike around the
    GC~\cite{Gondolo:1999ef, Fields:2014pia, Sadeghian:2013laa,
Ferrer:2017xwm}) and therefore improve the bounds significantly.

%---------------------------------------------------------------------
\paragraph{Testing EP with pulsars.}
%---------------------------------------------------------------------

The possibility to test the (strong) EP with binary pulsars was proposed by
\citet{Damour:1991rq}, utilizing the differential acceleration  from the
Galactic matter distribution on  the two components of a binary. The relative
acceleration reads, $\ddot{\bm{R}} = - {\cal G} M \hat{\bm{R}} / R^2 +
{\bm{A}}_{\rm PN} + {\bm{A}}_\eta$, where $\bm{R}$ is the relative separation,
${\cal G}$ denotes the effective gravitational constant, $M$ is the total mass
of the binary, $R \equiv \left| \bm{R} \right|$ and $\hat{\bm{R}} \equiv \bm{R}
/ R$.  In the above expression, ${\bm{A}}_{\rm PN} $ denotes the post-Newtonian
(PN) corrections~\cite{Damour:1991rq, Will:1993ns, Freire:2012nb}, and we
consider ${\bm{A}}_\eta$ as the EP-violating anomalous acceleration towards DM.
At leading order, $\bm{A}_\eta = \eta_{\rm DM}^{({\rm NS, WD})} \bm{a}_{\rm
DM}$ for a neutron star (NS) -- white dwarf (WD) binary~\cite{Damour:1991rq,
Adelberger:2009zz, Freire:2012nb}. It is better to view the {\it apparent} EP
violation arising from a new long-range interaction (namely fifth force)
between DM and ordinary matter~\cite{Fischbach:1985tk, Wagner:2012ui}. In the
following we use GR for gravity, and ${\cal G}$ becomes the Newtonian
gravitational constant $G$.

We denote $\hat{\bm{a}}$ the unit vector directing from the center of the
binary towards periastron, and $\hat{\bm{k}}$ the one along orbital angular
momentum.  After averaging over an orbit, the secular changes on the orbital
elements, introduced by the relative acceleration, are summarized as,
$\left\langle {\rmd P_b}/{\rmd t} \right\rangle = 0$, $\left\langle {\rmd
\bm{e}}/{\rmd t} \right\rangle = \bm{f} \times \bm{l} + \dot\omega_{\rm PN}
\hat{\bm{k}} \times \bm{e}$, and $\left\langle {\rmd \bm{l}}/{\rmd t}
\right\rangle = \bm{f} \times \bm{e}$. We have introduced $\bm{e} \equiv e
\hat{\bm{a}}$, $\bm{l} \equiv \sqrt{1-e^2} \hat{\bm{k}}$, $\bm{f} \equiv
\frac{3}{2} {\cal V}_{\rm O}^{-1} \bm{A}_\eta$~\cite{Damour:1991rq,
Freire:2012nb}, with $P_b$ the orbital period, $e$ the orbital eccentricity,
and ${\cal V}_{\rm O} \equiv \left( 2\pi G M/P_b \right)^{1/3}$. At first PN
order, the periastron advance rate reads, $\dot\omega_{\rm PN} = 6\pi \left(
{\cal V}_{\rm O} /c \right)^2 / \left[ P_b\left( 1-e^2 \right) \right]$.
Integrating the above differential equations gives the orbital dynamics, which
is concisely summarized as $ \bm{e}(t) = \bm{e}_{\rm PN} (t) + \bm{e}_\eta
$~\cite{Damour:1991rq}.  Graphically, the evolution of the orbital eccentricity
vector $\bm{e}(t)$ has two components, (i) a general-relativistically
precessing $\bm{e}_{\rm PN}(t)$ with a rate $\dot\omega_{\rm PN}$, and (ii) a
constant ``forced'' eccentricity, $\bm{e}_\eta \equiv \bm{f}_\perp
\dot\omega_{\rm PN}^{-1}$ with $\bm{f}_\perp$ the projection of $\bm{f}$ on the
orbital plane. This was used extensively to constrain EP with binary pulsars
under a probabilistic assumption on unknown angles~\cite{Damour:1991rq,
Wex:1999bx, Stairs:2005hu, Wex:2014nva, Zhu:2015mdo}.

Recently, a direct test, which evades the probabilistic assumption, was
proposed out of this framework~\cite{Freire:2012nb}.  It uses the time
derivatives of the orbital eccentricity vector, $\dot{\bm{e}}$, and of the
orbital inclination; the latter causes a nonzero time derivative of the
projected semimajor axis, $\dot x$.  These parameters are directly fitted from
the time-of-arrival data~\cite{Lorimer:2005}. The first implementation of the
idea, based on
pulsar timing data from EPTA and NANOGrav, was achieved by \citet{Zhu:2018etc}
with PSR~J1713+0747.  Using $\dot{\bm{e}}$ of PSR~J1713+0747~\cite{Zhu:2018etc}
and the acceleration from DM at its location~\cite{2017MNRAS.465...76M}, we
obtain $\left| \eta_{\rm DM}^{({\rm NS,WD})} \right| < 0.004$ at 95\% C.L..

%---------------------------------------------------------------------
\paragraph{Non-gravitational forces between DM and ordinary matter.}
%---------------------------------------------------------------------

%---
\begin{figure}
  \includegraphics[width=9cm]{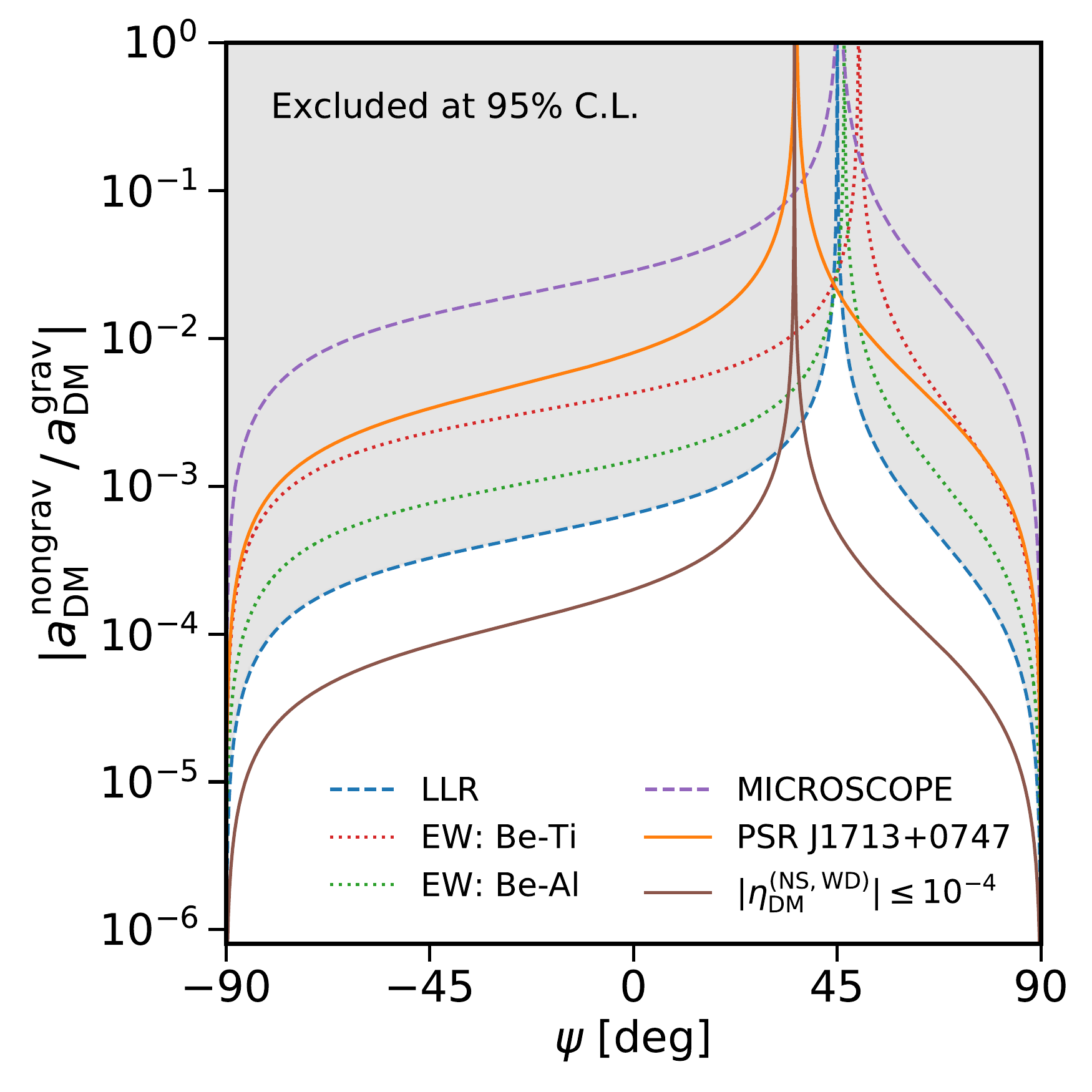}
  \caption{\label{fig:ng:proton} The $95\%$ C.L. limits on $\left|
      a^{\rm nongrav}_{\rm DM} / a^{\rm grav}_{\rm DM} \right|$ for
      neutral hydrogen, from LLR~\cite{1995A&A...293L..73N,
      Anderson:1995df, Williams:2005rv}, E\"ot-Wash (EW)
      experiments~\cite{Wagner:2012ui},
      MICROSCOPE~\cite{Touboul:2017grn}, and
      PSR~J1713+0747~\cite{Zhu:2018etc}. The expected limit from a
      {\it hypothetical} NS-WD system with a $1.4\,{\rm M}_\odot$ NS
      that constrains $\left| \eta_{\rm DM}^{({\rm NS,WD})} \right|
  \leq 10^{-4}$ at 95\% C.L. is also plotted.}
\end{figure}
%---

%---
\begin{table}
  \centering 
  \caption{Material sensitivities for different
  objects~\cite{Adelberger:2009zz}. For NSs, the gravitational binding energy
  is assumed to be proportional to mass~\cite{Damour:1992we, Zhu:2018etc}, and
  the composition is dominated by neutrons.  \label{tab:composition}}
  %---
  \begin{tabular}{p{2.5cm}p{1.5cm}p{1.5cm}p{1.5cm}}
  \hline\hline 
   & $Z/\mu$ & $N/\mu$ & $B/\mu$ \\
  \hline
  Be & 0.44384 & 0.55480 & 0.99865 \\
  Al & 0.48181 & 0.51887 & 1.00068 \\
  Ti & 0.45961 & 0.54147 & 1.00108 \\
  Pt & 0.39984 & 0.60034 & 1.00018 \\
  Earth & 0.49 & 0.51 & 1.00 \\
  Moon & 0.50 & 0.50 & 1.00 \\
  NS ($1.33\,M_\odot$) & $\simeq 0$ & $\simeq 1.19$ & $\simeq 1.19$ \\
  WD ($0.290\,M_\odot$) & $\simeq 0.5$ & $\simeq 0.5$ & $\simeq 1.0$ \\
  \hline
  \end{tabular}
\end{table}
%---

To interpret the result from pulsar timing, we will adopt the generic framework
widely used in testing EP~\cite{Stubbs:1993xk, Smith:1993un, Adelberger:2009zz,
Wagner:2012ui}. In quantum field theory, scalar or vector boson exchange
introduces a spin-independent potential between a test mass, $A$, and the
attractor (here the DM)~\cite{Fischbach:1985tk, Adelberger:2009zz,
Wagner:2012ui}, $V(r) = \mp g_5^2 q_5^{(A)} q_5^{\rm DM}e^{-r/ \lambda} / 4\pi
r $, where $g_5$ is the coupling constant, $q_5$ is the (dimensionless) charge,
and the upper (lower) sign is for scalar (vector) boson. From the potential one
has~\cite{Wagner:2012ui},
%---
\begin{equation}\label{eq:delta:eta}
  \eta_{\rm DM}^{({\rm A,B})} = \pm \frac{g_5^2}{4\pi G u^2} \frac{q^{\rm
  DM}_5}{\mu_{\rm DM}} \left[ \frac{q^{(A)}_5}{\mu_A} -
  \frac{q^{(B)}_5}{\mu_B}\right] \left( 1 + \frac{r}{\lambda} \right)
  e^{-r/\lambda} \,,
\end{equation}
%---
where $\left(q_5/\mu\right)$ is an object's charge per atomic mass unit, $u$.
Hereafter we will assume $\lambda \gg {\cal O}\left( 10\,{\rm kpc} \right)$, or
equivalently $m \ll 10^{-27}\,{\rm eV}/c^2$ for the mass of the boson field.

For test masses composed of ordinary matters ($p$, $n$, $e$), we parameterize
the charge $\left(q_5/\mu\right) = \left(Z/\mu\right) \cos\psi +
\left(N/\mu\right) \sin\psi$~\cite{Wagner:2012ui} with the mixing angle
satisfying $\tan \psi \equiv q_5^{(n)} / \left[ q_5^{(p)} + q_5^{(e)} \right]$.
This is the most general expression for vector charge, and a reasonable
tree-level approximation for scalar charge~\cite{Stubbs:1993xk,
Adelberger:2009zz}.  Notice that in Eq.~(\ref{eq:delta:eta}) the masses are
reduced according to objects' (negative) binding energies.  For ordinary
bodies, one has $\left(B/\mu\right) \equiv \left(Z/\mu\right) +
\left(N/\mu\right) = 1 + {\cal O}\left( 10^{-3} \right)$; see
Table~\ref{tab:composition}. For NSs, due to their significant gravitational
binding energy, $\left(B/\mu\right) = 1 + {\cal O}\left( 10^{-1} \right)$. 

In Figure~\ref{fig:ng:proton}, we plot the constraints on the ratio of
non-gravitational acceleration of neutral hydrogen to the total
acceleration towards the Galactic DM, $a^{\rm nongrav}_{\rm DM} /
a^{\rm grav}_{\rm DM}$, as a function of the (theory-dependent) mixing
angle.  Although PSR~J1713+0747 only limits $\left| \eta_{\rm
DM}^{({\rm NS,WD})} \right| \lesssim 0.004$, the vast material
difference between the NS and the WD boosts its constraint
significantly. We have updated an underestimation of the Galactic DM
acceleration from $5\times10^{-11}\,{\rm
m\,s}^{-2}$~\cite{Stubbs:1993xk, Smith:1993un, Wagner:2012ui} to
$9.2\times10^{-11}\,{\rm m\,s}^{-2}$~\cite{2017MNRAS.465...76M}, thus
tightening the limits in \citet{Wagner:2012ui} even further. As we can
see, because NSs' $\left( B/\mu \right)$ significantly deviates from
unity, the unconstrained region differs from $\psi \simeq 45^\circ$
for PSR~J1713+0747, and the limit around $\psi \simeq 45^\circ$ is
given by this binary. In the following, we discuss how the proposed
test will improve in future.

%---------------------------------------------------------------------
\paragraph{DM spikes around the GC.}
%---------------------------------------------------------------------

As stressed by~\citet{Hui:2016ltb}, DM models have seldom been successfully
examined by observations at scales $\lesssim 10\,$kpc, due to the complication
of baryonic physics and unknown DM properties.  Nevertheless, well motivated
models exist for the DM distribution around the GC. We consider generalized
Navarro-Frenk-White (gNFW) profile~\cite{Navarro:1996gj} augmented with DM
spikes around the supermassive black hole (BH) in the GC,
Sgr~A$^*$~\cite{Gondolo:1999ef, Vasiliev:2007vh, Fields:2014pia,
Sadeghian:2013laa, Ferrer:2017xwm, Shapiro:2016ypb}.

NFW profile is a common approximation to the density profile found in DM-only
cosmological simulations~\cite{Navarro:1996gj}. Here we consider a generalized
form,
%---
\begin{equation}\label{eq:gNFW}
  \rho_{\rm gNFW}(r) = \frac{\rho_0}{(r/R_{\rm s})^\gamma (1+r/R_{\rm
  s})^{3-\gamma}} \,,
\end{equation}
%---
where $R_{\rm s} = 20\,$kpc, and $\rho_0$ is fixed by requiring $\rho_{\rm
gNFW} = 0.4\,{\rm GeV\,cm}^{-3}$ at the location of our Solar System ($r \simeq
8\,{\rm kpc}$).  The canonical NFW profile used $\gamma=1$; it fits well the
outer Galactic halo (see {\it e.g.} \citet{2017MNRAS.465...76M}).  We are
mostly interested in the inner region of the halo, and will consider $\gamma
\in [1.0, 1.4]$, motivated by numerical simulations ({\it e.g.} $\gamma \simeq
1.24$ in \citet{Diemand:2008in}) and Fermi-LAT $\gamma$-ray observations ({\it
e.g.} $\gamma \simeq 1.26$ in \citet{Daylan:2014rsa}).

\citet{Gondolo:1999ef} pointed out that, the gNFW model cannot give accurate
description for the inner sub-parsec region close to the GC.  In response to
the adiabatic growth of Sgr~A$^*$, a DM spike with $\rho_{\rm sp}(r) \propto
r^{-\gamma_{\rm sp}}$ will form, with $\gamma_{\rm sp} = \left( 9-2\gamma
\right)/\left( 4-\gamma \right)$ for collisionless DM. It forms inside the
radius of gravitational influence $R_{\rm h} \equiv GM_\bullet/v_0^2 \simeq
1.7\,{\rm pc}$ of Sgr~A$^*$, where $M_\bullet \simeq 4\times10^6 \,M_\odot$ is
the mass of BH and $v_0 \simeq 105\,{\rm km\,s}^{-1}$~\cite{Gultekin:2009qn} is
the one-dimensional velocity dispersion of DM in the halo outside the spike.
Including GR effects~\cite{Sadeghian:2013laa} and the rotation of
BH~\cite{Ferrer:2017xwm} will further enhance the spike.  Nevertheless, the
maximum density of the spike is limited by the annihilation cross-section of DM
particles~\cite{Vasiliev:2007vh}, producing $\rho_{\rm in}(r) \propto
r^{-\gamma_{\rm in}}$ ($\gamma_{\rm in} \simeq 0.5$ for $s$-wave annihilation
and $\gamma_{\rm in} \simeq 0.34$ for $p$-wave
annihilation~\cite{Shapiro:2016ypb}). Such a weak annihilation cusp happens
inside $R_{\rm in}\sim{\rm mpc}$ where the density reaches the ``annihilation
plateau'' $\rho_{\rm ann} \simeq 1.7\times10^8 \, M_\odot\,{\rm pc}^{-3}$. 

Taking the above results into consideration, we use a DM density profile,
%---
\begin{equation}
  \rho_{\rm DM}(r) = 
  \begin{dcases}
    \frac{\rho_{\rm sp} (r) \rho_{\rm in}(r)}{ \rho_{\rm
    sp} (r) + \rho_{\rm in}(r)}  \,, &  4GM_\bullet / c^2 \leq r <
    R_{\rm sp} \\
    \rho_{\rm gNFW}(r) \,, & r \geq R_{\rm sp}
  \end{dcases}
  \label{eq:dm:profile}
\end{equation}
%---
where three different values for $R_{\rm sp}$ ($= \frac{1}{5}R_{\rm h}$,
$R_{\rm h}$, $5R_{\rm h}$) are adopted for illustrating purposes. Normalization
factors for $\rho_{\rm sp}(r)$ and $\rho_{\rm in}(r)$ are obtained by
continuity.

%---
\begin{figure}
  \includegraphics[width=9cm]{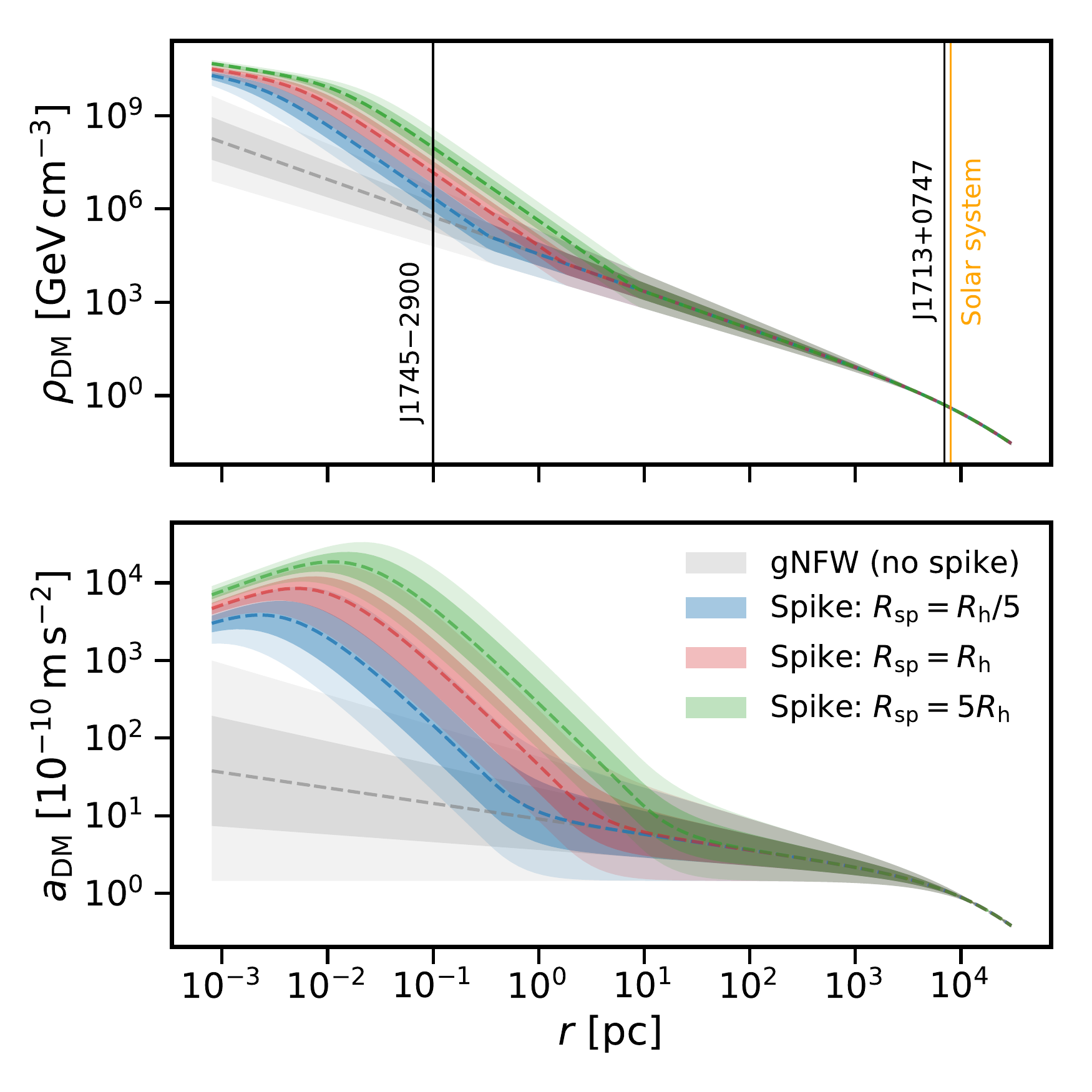}
  \caption{\label{fig:dm:acc} {\it (Upper)} DM density as a function of the
      distance to the GC.  {\it (Lower)} Acceleration produced by the DM inside
      radius $r$. The lighter and darker shadowed regions enclose $\gamma \in
      [1.0, 1.4]$ and $\gamma \in [1.1, 1.3]$ respectively, while the dashed
      lines give $\gamma = 1.2$. In the upper panel, locations of
  PSR~J1745$-$2900~\cite{Eatough:2013nva}, PSR~J1713+0747~\cite{Zhu:2018etc},
  and the Solar System are indicated.}
\end{figure}
%---

DM density profiles at different radii are given in the upper panel of
Figure~\ref{fig:dm:acc} for ``gNFW'' indices $\gamma \in [1.0, 1.4]$. The
steepening of the DM spike (relative to the ``gNFW'' profile) happens at
$r\sim10\,{\rm pc}$, and its flattening happens at $r\sim10^{-2}\,{\rm pc}$
($R_{\rm in}=2.7\,{\rm mpc}$, $5.9\,{\rm mpc}$, $13\,{\rm mpc}$ for the three
$R_{\rm sp}$'s when $\gamma=1.2$)~\cite{Fields:2014pia}. In the lower panel we
give the acceleration produced solely by the DM, $a_{\rm DM}(r) \equiv G \left.
\int^r 4\pi r'^{2} \rho_{\rm DM}(r') {\rm d} r' \right/ r^2$. One can see
that, $a_{\rm DM}(r) / a_{\rm DM}^\odot \leq 1.1$ at the location of
PSR~J1713+0747~\cite{Zhu:2018etc}, where $a_{\rm DM}^\odot$ is the DM
acceleration at the Solar System. However at the location of the magnetar
PSR~J1745$-$2900 ($r\sim0.1\,{\rm pc}$)~\cite{Eatough:2013nva}, depending
on the value of $\gamma$, this quantity can be as high as 23--870 for
$R_{\rm sp}=\frac{1}{5}R_{\rm h}$, 180--3600 for $R_{\rm sp}=R_{\rm h}$,
and 1400--14000 for $R_{\rm sp}=5R_{\rm h}$.  This factor will be the
\textit{gain} in the driving force to test the universality of free fall
(UFF) towards DM if a binary pulsar is found there.

%---------------------------------------------------------------------
\paragraph{Pulsar surveys towards the GC.}
%---------------------------------------------------------------------

The magnetar PSR~J1745$-$2900~\cite{Eatough:2013nva} is already within the most
interesting region, but unfortunately it is not in a binary. The closest
binaries known so far are PSRs J1755$-$2550 and J1759$-$24, with radial
distances from the GC of about 2\,kpc, although the exact distances are still
highly uncertain \cite{Ng:2015zza, Ng:2018lkc}. Future radio surveys are likely
to overcome existing selection effects and promise to find binaries in much
closer proximity. In particular the SKA has the capability of finding nearly
all radio pulsars beamed towards the Earth \cite{Kramer:2004hd}, including
those pulsars near the GC~\cite{Eatough:2015jka}. Already the first phase,
SKA1, should find about 10000 pulsars \cite{keane:2015} in the Galaxy, about
$10$\% of which can be expected to be in binaries, based on the currently known
population. In order to probe the GC region, high-frequency surveys may be
needed to overcome the scattering of the radio pulses in the turbulent
interstellar medium \cite{Lorimer:2005}, but such surveys are ongoing already
and further are planned \cite{Goddi:2017pfy}. Constraints on the pulsar
population from observations at multi-wavelengths around the
GC~\cite{Wharton:2011dv} suggest that the inner parsec of the Galaxy could
harbor as many as $\sim10^3$ active radio pulsars that are beaming toward the
Earth. Those pulsars should include a number of suitable binaries, and
simulations show that even a few PSR-BH systems should be present in the
central parsec today \cite{fl11}, which would be prime
targets for the studies suggested here.

%---------------------------------------------------------------------
\paragraph{Discussions.}
\label{sec:disc}
%---------------------------------------------------------------------

In this {\it Letter} we propose to use radio timing of binary pulsars to
constrain non-gravitational forces between DM and ordinary matter that will
appear as an {\it apparent} violation of EP towards DM. As we can see in
Figure~\ref{fig:ng:proton}, the current limit on UFF from
PSR~J1713+0747~\cite{Zhu:2018etc} is already providing important improvement
over currently best limits~\cite{1995A&A...293L..73N, Wagner:2012ui}.  The test
with pulsars has unique advantages over other tests, that we will recapitulate
and further elaborate below.
%---
\begin{itemize}
  \item {\it Driving force.}
    Driving force sets an important reference in testing EP. At the site of the
    E\"ot-Wash laboratory in Seattle, the driving forces from the Earth, the
    Sun, and DM, are $1.68\times10^{-2} \, {\rm m\,s}^{-2}$, $5.9\times10^{-3}
    \, {\rm m\,s}^{-2}$, and $9.2\times10^{-11} \, {\rm m\,s}^{-2}$,
    respectively, thus the 1\,$\sigma$ limits (from the Be-Ti pair) are
    $\left|\eta_\oplus\right| \lesssim 2\times10^{-13}$,
    $\left|\eta_\odot\right| \lesssim 5\times10^{-13}$, and $\left|\eta_{\rm
    DM}\right| \lesssim 3\times10^{-5}$~\cite{Wagner:2012ui}. To test
    $\eta_\oplus$, MICROSCOPE gains a factor of 500 in the driving force by
    putting the experiment in space~\cite{Touboul:2017grn}. However, it does
    not have such a gain when the attractor is DM. For the same reason, the
    triple pulsar~\cite{Ransom:2014xla,Shao:2016ubu}, while gaining a factor of
    ${\cal O}\left( 10^7 \right)$ in driving force to test the strong EP,
    cannot probe the UFF towards DM at a comparable level.  As shown in
    Figure~\ref{fig:dm:acc}, if future surveys find suitable binary pulsars
    within ${\cal O}\left( 10\,{\rm pc} \right)$ of the GC, the driving force
    can easily be enhanced by orders of magnitude.
  \item {\it Measurement precision.}
    \citet{Freire:2012nb} showed that uncertainties in $\dot e$ and $\dot x$
    (denoted as $\delta \dot e$ and $\delta \dot x$) scale as $\delta \dot e
    \simeq 8.0 \, \delta t / x \sqrt{\bar N T^3}$ and $\delta \dot x \simeq 5.3
    \, \delta t / \sqrt{\bar N T^3}$, where $\bar N$ is the average number of
    TOAs per unit time, $\delta t$ is the rms of TOA residuals, and $T$ is the
    observing baseline. Even with current pulsars, longer observations will
    improve the test as $T^{-3/2}$, and future telescopes, like
    FAST~\cite{Nan:2011um} and SKA~\cite{Kramer:2004hd, Shao:2014wja}, will be
    able to improve the timing precision $\delta t$ significantly. Therefore,
    the proposed test will improve continuously. In addition, it was shown that
    the $\dot{\bm{e}}$-test is a clean test, not being contaminated by external
    effects~\cite{Freire:2012nb}.
  \item {\it Material sensitivity.} 
    NSs are unique in the sense that they contain a dominant portion of
    neutron-rich materials. This gains a factor of ${\cal O}\left( 10^2
    \right)$ for most $\psi$'s when interpreting the $\eta_{\rm DM}$ limit in
    Figure~\ref{fig:ng:proton}. Depending on the equation of state, NSs might
    contain exotic excitations like pions and kaons. It would allow to test
    couplings of DM with these degrees of freedom that is inaccessible with
    other alike experiments.
  \item {\it Binding energy.} 
      As ordinary matter has $\left( B/\mu \right) \simeq 1$, any
	  individual experiment will have an infinite peak around
	  $\psi\simeq 45^\circ$ in Figure~\ref{fig:ng:proton}.  This
	  infinite peak can be removed by combining results from two
	  or more different test-body pairs.  Due to the significant
	  gravitational binding energy of NSs, the peak is shifted
	  towards smaller $\psi \simeq \tan^{-1} \left( 1-2\epsilon
	  \right)$, where $\epsilon$ is the (absolute value of)
	  fractional gravitational binding energy.  The combination of
	  our limits from pulsar timing with existing experiments
	  gives an improved constraint in the region around
	  $\psi\simeq 45^\circ$.  Future results from other pulsar
	  binaries close to the GC have the possibility of making
      substantial improvements over most of the range of the $\psi$
  parameter (see Figure~\ref{fig:ng:proton}).
\end{itemize}
%---
It is pleasing to see that pulsar timing naturally possesses all the advantages
mentioned above to boost its test of UFF towards DM.  Although a binary pulsar
at the GC will most certainly not have the same timing precision as
PSR~J1713+0747, due to the boost in the driving force by orders of magnitude,
it might still allow for a limit of $\left| \eta_{\rm DM}^{({\rm NS,WD})}
\right| \lesssim 10^{-4}$.  In Figure~\ref{fig:ng:proton} we plot the
corresponding constraint, that will exclude non-gravitational force between DM
and neutral hydrogen at 1\textperthousand{} level for any mixing angle.

%---------------------------------------------------------------------
\acknowledgments
%---------------------------------------------------------------------

We are grateful to Ralph Eatough, Paulo Freire, and Anna Nobili for
discussions, and the anonymous referees for comments that improved the
manuscript.  We acknowledge financial support by the European
Research Council (ERC) for the ERC Synergy Grant BlackHoleCam under
contract no. 610058.

\end{document}